\documentclass[doublecol,nopacs]{epl2} 
\usepackage{graphicx,amssymb,amsfonts,amsmath,float,color}

\renewcommand{\pacs}[2]{}
\makeatletter
\renewcommand{\epl@pacsmissing}{}
\makeatother
\newcommand{\beq}{\begin{equation}}
\newcommand{\eeq}{\end{equation}}
\newcommand{\bea}{\begin{eqnarray}}
\newcommand{\eea}{\end{eqnarray}}
\newcommand{\no}{\nonumber}
\newcommand{\ee}{entanglement entropy}
\newcommand{\ka}{\kappa}
\newcommand{\bka}{\bar{\kappa}}
\newcommand{\gap}{
\begin{figure}
 \includegraphics[width=8cm,clip]{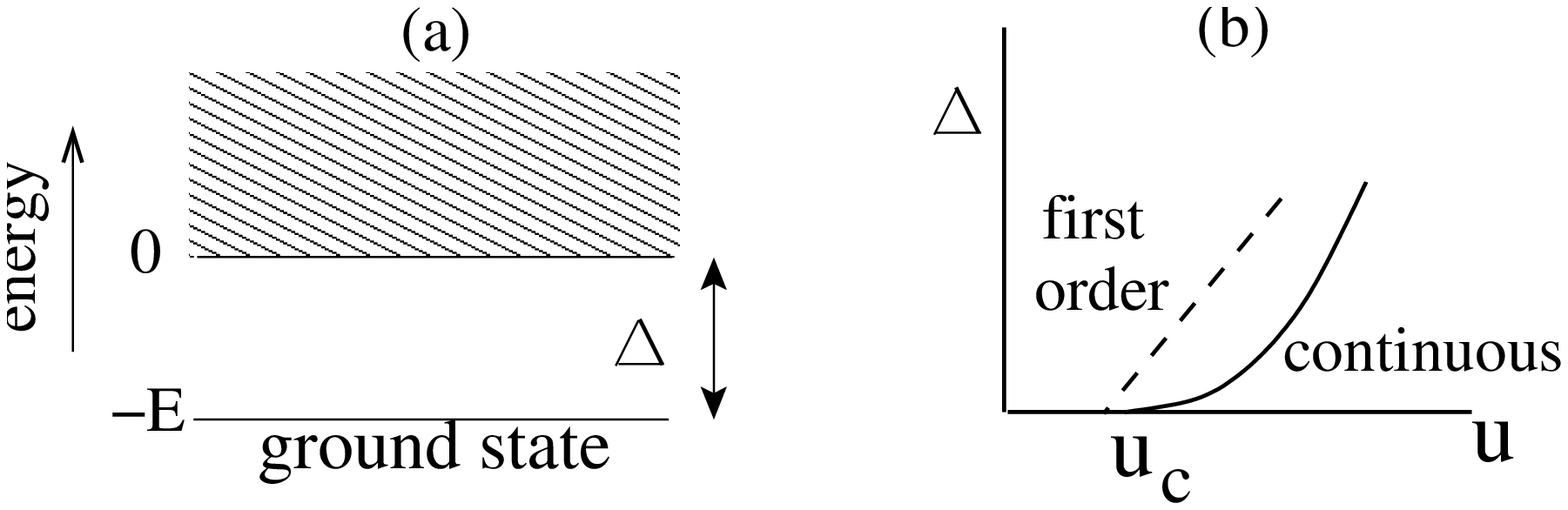}
\caption{{(a) Gap $\Delta$ in the energy  spectrum. The shaded region
  is the continuum of energy.
(b) The graph shows how energy gap goes to zero. The continuous line
describes a  second order or continuous
transition (critical) and the dashed line shows the first order
transition. The two are distinguished by the behaviour of the slope at
$u=u_c$.}}
\label{fig:gap}
\end{figure}
}
\newcommand{\bubble}{
\begin{figure}
 \begin{center}
 \includegraphics[scale=0.35,clip]{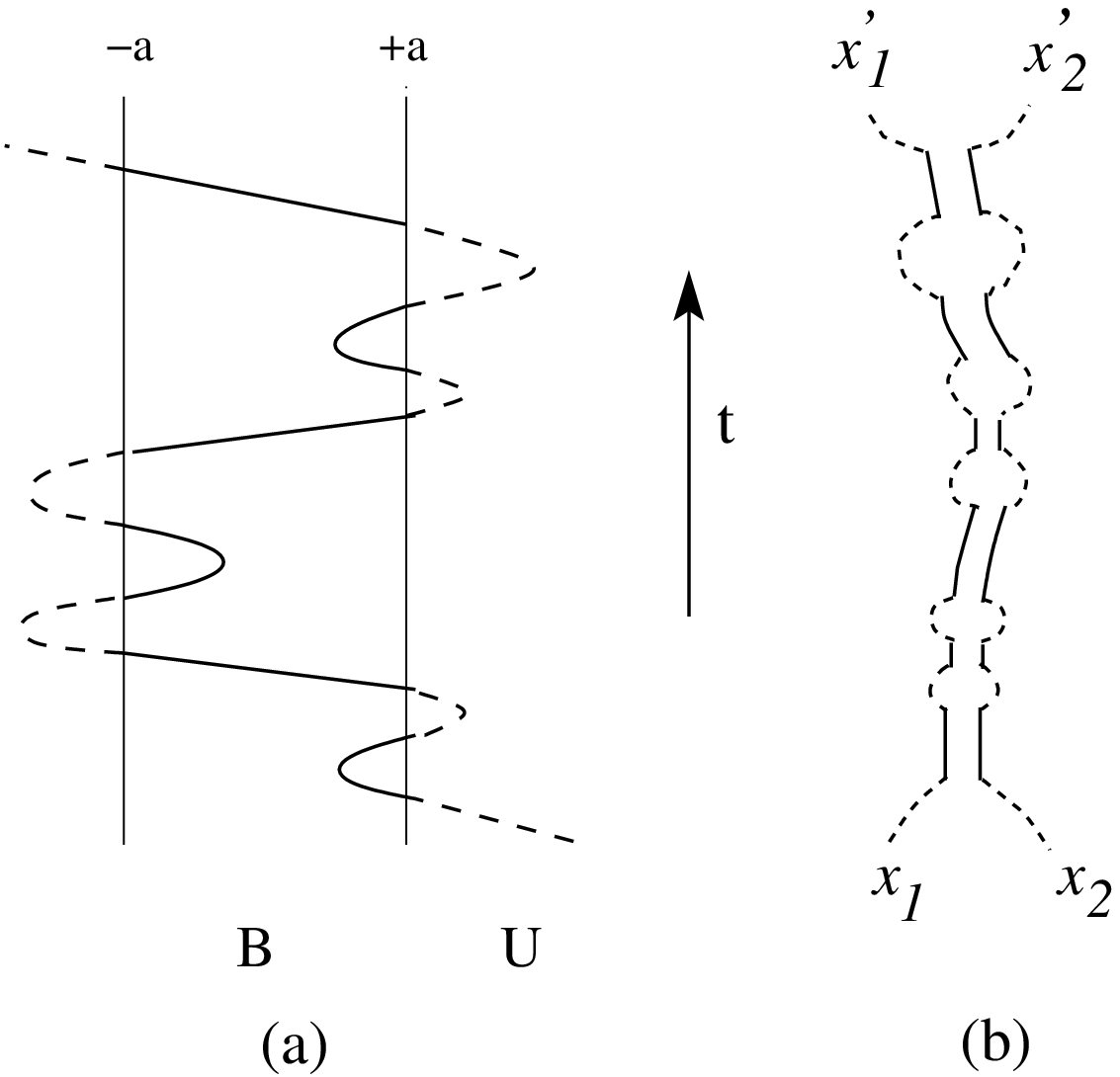}
\end{center}
\caption{{Path integral representation in the $x$-$t$ plane.  (a) A
  relative coordinate path for two particles in one-dimension.  The
  solid portions represent the classical bound state, i.e., inside the
  well (B), and the dashed portions represent the unbound (U) state in
  the classically forbidden region.  (b) Corresponding path
  representation of two quantum particles with time, though
  intersections of paths are not shown explicitly.  It is also a
  configuration of two classical Gaussian polymers interacting at the
  same contour length as for DNA base pairing, the $t$-axis
  representing the contour length ($z$) of the polymers. The dotted
  lines are the melted bubbles whose partition functions are
  characterized by the reunion exponent $\Psi$. This description 
holds for any general $d$.} }
\label{fig:bubb}
\end{figure}
}

\newcommand{\sfit}{%
\begin{figure}[h]
\begin{center}
 \includegraphics[scale=0.5,clip]{3d_fit.eps}
\end{center}
\caption{{Plot of $S$ vs. $\ln\,\ka$ with $a=1$. The circles are the numerical 
values and the straight line is the  predicted line
$S=3\ln\,\ka+7.06$, Eq. \ref{eq:threed}.}}
\label{fig:1}
\end{figure}
}

\title{Entanglement entropy of a quantum unbinding transition\\ and entropy of DNA}
\shorttitle{Negative entanglement entropy} 

\author{Poulomi Sadhukhan\inst{1}\thanks{poulomi@iopb.res.in} \and  
Somendra M. Bhattacharjee\inst{1}\thanks{somen@iopb.res.in}}
\shortauthor{Poulomi Sadhukhan and Somendra M. Bhattacharjee}

\institute{                    
  \inst{1}Institute of Physics, Bhubaneswar 751 005, India.
}

\abstract{Two significant consequences of quantum fluctuations are 
  entanglement and criticality.  Entangled states may not be critical 
  but a critical state shows signatures of universality in 
  entanglement.  A surprising result found here is that the 
  entanglement entropy may become arbitrarily large and negative near 
  the dissociation of a bound pair of quantum particles.  Although 
  apparently counter-intuitive, it is shown to be consistent and 
  essential for the phase transition, by mapping to a classical 
  problem of DNA melting. We associate the entanglement 
  entropy to a subextensive part of the entropy of DNA bubbles, which 
  is responsible for melting.  The absence of any extensivity 
  requirement in time makes this negative entropy an inevitable 
  consequence of quantum mechanics in continuum.  Our results 
  encompass quantum critical points and first-order transitions in 
  general dimensions.}

\begin{document}

\maketitle

\section{Introduction}
Quantum entanglement\cite{horo, amico,indrani,kitaev} is a fundamental
feature of quantum mechanics, which says that performing a local
measurement may instantaneously affect the outcome of local
measurements far away.  There is another feature of quantum mechanics
where the zero-point fluctuations in the ground state may coherently
add up to produce long-range correlations of local observables.  This
happens at quantum critical points (QCP), a point where the spectrum
becomes gapless, obtained by tuning the parameters of the Hamiltonian.
In both cases, a pure state cannot be written as the product of the
wave-functions of the two distant parts, though states may be
entangled without being critical.  The ground-state energy may be
non-analytic through a quantum phase transition (QPT) or through a
quantum critical point.  The wave function encodes not only this
non-analyticity but also the special quantum correlations or quantum
entanglement intrinsic to the state.  

At or near a QCP, the signatures of its universality can therefore be
found in the entanglement, a common measure of which is the von
Neumann entropy ($S$)\cite{indrani,kitaev,osterloh,jphysa,kopp,igloi}.  The exact
results of this paper show that for a class of critical points, viz.,
the dissociation of a pair of particles in the unitarity limit of
infinite scattering length, there is the possibility of a negatively
diverging $S$.  Although counter-intuitive, this is not an artifact.
Analogous situation occurs in statistical mechanics for Gibbs entropy
in canonical ensemble for a gapless spectrum. As discussed below, the
problem in hand involves a gapless entanglement spectrum. The usual
proof of the positivity of entanglement entropy is not applicable in
case of continuous eigenvalues of the reduced density matrix. The
negative entropy is essential for the criticality itself. Its
importance is brought out via the mapping of the quantum problem to
the equivalent classical statistical mechanical problem, the melting
of a double-stranded DNA\cite{fish, smb,fisher,smlong}.

\section{Entanglement entropy} 
Recall the problem of a quantum particle of mass $m$ in a three 
dimensional spherical potential well, 
\begin{eqnarray} 
  \label{eq:1} 
  V({\bf r}) &=& -V_0 \qquad {\rm for}\ r<a,\nonumber\\ 
                 &=& \ \ 0 \qquad\ \ {\rm for} \ r>a, 
\end{eqnarray} 
where $r$ is the radial coordinate, $a$ and $V_0$ are the width and
the depth of the potential well.  What is special is that $V_0>0$ does
not guarantee the existence of a bound state, unlike in one or two
dimensions, or in classical mechanics.  No bound state exists for
$u<u_c$ where $u=2 m V_0 a^2/\hbar^2$ is the dimensionless parameter
for the potential and $u_c$ corresponds to a critical value of $u$.
For simplicity, we take $u\approx u_c$ so that there is only one bound
state. In this situation energy $\mid\!\! E\!\!\mid$ itself is the gap
in the spectrum.  If we tune $u$ to get a state with zero energy
($E=0$), then at that energy in $d=3$ the wave function $\varphi(r) \sim
1/r$ which is like a non-normalizable critical state.  Like a bound
state the probability density does decay to zero but like an unbound
state it is not normalizable.  In higher dimensions, the condition for
a minimal strength of the potential for a bound state remains true,
but the state corresponding to $E=0$ becomes normalizable as it should
be for a bound state.  So we see that this bound to unbound transition
for a potential well has different nature in different dimensions.  In
general, {\it(i)} for $d\leq 2$ there is no such transition as $E=0$
requires $V_0=0$, though there are remnants of the transition as
$V_0\to 0$, {\it (ii)} for $2<d<4$, the transition is continuous
(critical) --- the bound state becomes unbound through a
non-normalizable critical state as we change $u$, and, {\it (iii)} for
$d>4$, the bound state remains normalizable up to and including $E=0$,
and becomes unbound as $u$ is decreased further, thus making the
transition first-order. This depicts a QPT
and the case of a potential well gives a simple example of a quantum
critical point for $2\!<\!d\!<\!4$ with diverging length scales.

\gap

The ground state energy, for $u$ close to $u_c$, is the gap
  $\Delta$ in the spectrum. A quantum phase transition is
  characterized by a vanishing gap.  A  discontinuity of the
  first derivative $d\Delta/du$ signals  a first order transition,
  otherwise it is critical or continuous, as
  shown in Fig.  \ref{fig:gap}.  One may define characteristic  time
  and length  scales 
  \begin{equation}
    \label{eq:14}
    \xi_{\parallel} = {\hbar}{\Delta}^{-1}, \ {\rm and} \ \xi_{\perp}={\hbar}/{\sqrt{2m\Delta}},
  \end{equation}
both of which  diverge as $\Delta\to 0$, with
  $\xi_{\parallel}\sim \xi_{\perp}^z$, $z$ ($=2$ in this case) being
  the dynamic exponent.  One may compare with the classical ground state to see the
  importance of quantum (zero-point) fluctuations and the importance
  of  time or dynamics in quantum phase transitions.  A path-integral
  interpretation of these scales, useful for the DNA mapping, is given
  below.    
 
Let us now consider the ground state of two dissimilar particles 
interacting via a central potential $V\!(\mid\!\! {\bf r}_1\!-\!{\bf 
  r}_2\!\!\mid)$ of the type of Eq. \ref{eq:1}, with ${\bf r}_i$ 
denoting the co-ordinate of the $i$-th particle. The existence of 
diverging length scales and scaling behavior around $u=u_c$ justifies 
the dissociation of 
the bound pair to be a QCP or a QPT depending on the dimensions 
they are in. 
The criticality is described by the exponents for the diverging length 
scales and the energy, as
\begin{equation}
\label{eq:expo}
|E| \sim \xi_{\parallel}^{-1}\sim \mid\! u-u_c\!\mid
^{\nu_{\parallel}},\ {\rm and}\  \xi_{\perp} \sim 
\mid u-u_c\mid^{-\nu_{\perp}},
\end{equation}
with
\bea
\label{eq:eng}
   \nu_{\parallel}=z\nu_{\perp}&=&{1}/{(\Psi -1)}, \ {\rm for}\
   1<\Psi \leq 2,\\
    &=&1 , \quad\quad\quad\quad {\rm for}\
   \Psi \geq 2,
\eea
which involve {\it (i)}  $z$ the dynamic exponent, and {\it (ii)} a universal 
exponent $\Psi$, known  as the  reunion exponent for  polymers \cite{fish,smb,fisher}.  
For the short range interaction problem, $\Psi=d/2$, as for random
walkers,  from which the
specialty of $d=4$ is apparent.

 \bubble

In a quantum bound state a particle can tunnel through the potential.
In a path integral approach the particle does a sizable excursion in
the classically forbidden region outside the interaction well, sooner
or later returning to the well (see Fig. \ref{fig:bubb}).  That the
two particles will eventually be close-by to form a bound state is the
source of entanglement while the excursions produce spreads of the
trajectories in space and time.  These spreads give the two relevant
length scales $\xi_{\parallel},\xi_{\perp}$.  The large width of the
bound state wave function near the QCP ensures the mutual influence of
the particles even if far away from each other ($r\gg a$) so that the
reduced density matrix for one particle still carries the signature of
the entanglement and the criticality.  For this bipartite system, we
are interested in the ``particle-partitioning
entanglement''\cite{haq}. This makes the von Neumann entropy a
valuable quantity for the transition which reads, 
{
\begin{equation} 
  \label{eq:2} 
  S=- {\rm Tr}\  \rho\  \ln \rho, 
\end{equation} 
 where $\rho$ is 
the reduced density matrix for the ground
state $|\psi\rangle$,
\begin{equation} 
  \label{eq:3} 
  \rho(\textbf{r}_1,\textbf{r}_1')= {\rm Tr}_2\ \varrho(1,2)=\int d^d
  \textbf{r}_2\ \langle\!\textbf{r}_1,\textbf{r}_2\!|\psi\rangle\langle\! \psi\!|\textbf{r}_1',\textbf{r}_2\rangle, 
\end{equation} 
obtained from the two particle density matrix
\hbox{$\varrho(1,2)=\mid\!\psi\rangle\langle\!\psi\!\mid$} by
integrating out (or tracing out) particle 2.  In Eq. \ref{eq:2}, we
shall introduce some pre-chosen length scale to make the argument of
log dimensionless.  If, with $m_i,\textbf{r}_i$ denoting the mass and
the position of the $i$th particle, the full ground state
wave-function (including the center of mass (CM)) is 
\beq
\label{eq:entdef}
\psi(\textbf{r}_1,\textbf{r}_2)=
\Phi\left(\frac{m_1\textbf{r}_1+m_2\textbf{r}_2}{m_1+m_2}\right)\,\varphi(\textbf{r}_1-\textbf{r}_2),
\eeq 
where $\Phi$ is CM wave function {(plane waves)} and
$\varphi$ is the wave function in relative coordinate (the relative
wave-function), then
\begin{equation}
  \label{eq:15}
  \rho(\textbf{r}_1,\textbf{r}_1')=\int d^d \textbf{r}_2\
\psi(\textbf{r}_1,\textbf{r}_2)\psi^*(\textbf{r}_1',\textbf{r}_2). 
\end{equation}
Although the center of mass and the relative parts are not entangled,
the two particles are entangled.  The lack of knowledge of the state
of one particle is the source of a nonzero entropy associated with the
reduced density matrix\cite{horo,amico,kitaev}.  
}

The translational invariance of the interaction guarantees that the
reduced density matrix $\rho({\bf r},{\bf r}^{\prime})\equiv \rho({\bf
  r - r^{\prime}})$ has $\exp(- i {\bf q}\cdot {\bf r})$ as the
eigenvector,
{ 
\bea 
  \int\!\! d^d\textbf{r}^{\prime} {\rho}(\textbf{r}-\textbf{r}^{\prime})
  e^{-i\textbf{q}\cdot\textbf{r}^{\prime}}\!\!\!\!  &=&
  \hat{\rho}(\textbf{q})\, e^{-i\textbf{q}\cdot\textbf{r}}, 
\eea 
}
{ with the eigenvalue 
  \begin{equation}
    \label{eq:17}
\hat{\rho}({\bf q})=\left|\phi\left({\bf  q}+\frac{{\textbf
        K}\mu}{m_2}\right)\right|^2,    
  \end{equation}
  $\textbf{K}$ being CM wave vector and ${\phi}({\bf q})$ the
  normalized momentum space wave function, the Fourier transform of
  the relative wave-function $\varphi({\textbf r})$ in Eq.
  \ref{eq:entdef}.}  Since the entropy involves an integral over the
whole range of ${\bf q}$, it is independent of the CM wave-vector, an
expected consequence of Galilean invariance.  Therefore, without any
loss of generality, we choose $\mid \textbf{K}\mid=0$.  The
eigenvalues constituting the ``entanglement spectrum'' can be written
in a scaling form
\begin{equation} 
  \label{eq:4} 
  \mid\! \phi({\bf q})\!\mid^2= \kappa^{-d} \ F({\bf q}/\kappa, a\kappa), 
\end{equation} 
where $\kappa^2= 2 \mu\! \mid\!\!
E\!\!\mid\!\!/\hbar^2=\xi_{\perp}^{-2}$, $\mu$ being the reduced mass.
Eq. \ref{eq:4} satisfies ${\rm Tr}\ {\hat{\rho}} =1$.  In the critical
regime (also called the ``unitarity limit''), $a\kappa \to 0$, if the
scaling function behaves smoothly, then
\begin{equation}
  \label{eq:16}
  F({\bf \tilde{q}},a\ka)\to  F({\bf \tilde{q}},0)\equiv f({\bf \tilde{q}}),
\quad (\tilde{q}\equiv q/\kappa) 
\end{equation}
which we find to be true for $d<4$.  For $d\geq 4$,  we find
that $F({\bf \tilde{q}},a\ka)$ for $a\ka \to 0$ behaves in a singular fashion as
\begin{equation}
  \label{eq:18}
  F({\bf x},y)\sim y^{d-4} f({\bf x}),
\end{equation}
so that the prefactor in Eq.  \ref{eq:4} becomes $\kappa^{-4}
a^{d-4}$.  Here $f$ represents a generic function. 
By using these limiting forms, we find the \ee \ to be
\begin{subequations}
\begin{eqnarray} 
  \label{eq:5} 
S= P \, \ln a\kappa + c_0,&\\ 
 P=\min(d,4),\ {\rm and} &
 c_0= - \int d^d x \ f(x) \ln f(x).
\end{eqnarray} 
\end{subequations}
{The last statement can be verified by direct computation of
  the momentum distribution function of the relative motion in
  $d$-dimensions.  There are further log-corrections at $d=2$ and
  $d=4$ which we do not discuss here.

To motivate Eq. \ref{eq:5} let us consider a few examples.  Consider
the quantum problem of two particles interacting via a delta-function
potential in one dimension: $ V({\bf x}) = -v_0\delta(x)$.  By using
the center of mass and the relative coordinate wave-function, we write
the wave function as 
\bea 
\psi(x_1,x_2)&=&C\
e^{iK\mu\left(\frac{x_1}{m_2}+\frac{x_2}{m_1}\right) }\ e^{-\kappa\mid
  x_1-x_2\mid} 
\eea 
which is translationally invariant.  Here $K$ is
the CM wave vector, $\kappa=\xi_{\perp}^{-1}$, and $C$ is the
normalization constant.  The reduced density matrix for particle 1 is
then 
\beq 
\rho(x,x^{\prime})= \frac{C^2}{\kappa}\
e^{-(iK\mu/m_2+\ka)|x^{\prime}-x|}\left[1+\kappa|x-x^{\prime}|
\right]
\eeq 
having eigenvalues (Eq. \ref{eq:17}) 
\bea
\hat{\rho}({\bf q})=
\frac{2}{\pi}\frac{1}{\kappa}\frac{1}{(1+\tilde{q}^2)^2},\
(K=0), 
\eea 
which is of the form
Eq. \ref{eq:16} with $f({\bf \tilde{q}})\sim (1+\tilde{q}^2)^{-2}$.  By introducing
an arbitrarily chosen well strength $\bar{v}$ or a scale
$a=\hbar^2/2\mu \bar{v}$ in Eq. \ref{eq:2}, the \ee \ is found to be
of the form of Eq. \ref{eq:5} with \beq
  \label{eq:oned}
P=1,\ {\rm and}\  c_0=\ln 8\pi -2. 
   \eeq 
For  $\kappa\to 0$, $\hat{\rho}({\bf q})\to \delta(q)$ with $S=0$.  There is a
difference between $\kappa\to 0$ and $\kappa=0$.

For a one-dimensional problem with the potential of Eq. \ref{eq:1},
one can go over to the delta function potential problem by taking
$a\rightarrow 0$ keeping $V_0a=v_0$ constant to get the same
$\ln\,\ka$ behaviour as in Eq. \ref{eq:oned}.

We then check for a
$3$-dimensional potential well, Eq. \ref{eq:1}.
The relative wave-function ($l=0$) for this potential is
\beq
\label{eq:3d}
\varphi ({\bf r}) =
\begin{cases}
  A\ \frac{\sin kr}{r}\quad\quad r<a \\
 B\ \frac{e^{-\ka r}}{r} \quad\quad r>a,
\end{cases}
\eeq 
with $k$ and $\ka$
as defined earlier and constants $A, B$ determined in the usual way of
continuity of the wave function and its derivative.  A direct Fourier
transformation of $\varphi({\bf r})$ has been used to numerically compute
the \ee.  To derive an analytical formula, we note that the dominant
contribution in $\hat{\rho}({\textbf q})$ in the limit
$a\ka\rightarrow 0$ comes from the outer part. In this approximation 
we get 
\beq 
\label{eq:dens}
\hat{\rho}({\textbf q}) = \frac{1}{\ka^3}\frac{1}{\pi^2}
\left(\frac{1}{1 + \tilde{q}^2}\right)^2 = \ka^{-3} f({\bf \tilde{q}}).
\eeq 
This $\hat{\rho}({\textbf q})$ satisfies the normalization
condition $\int d^3 q \ \hat{\rho}({\textbf q})=1$.  Thus for the $3$D
potential well interaction, the \ee \ is of the form of Eq. \ref{eq:5}
with \beq
\label{eq:threed}
P = 3, \ {\rm and}\   c_0=2 (1+\log (4 \pi ))\approx 7.06205.
\eeq

Exact numerical computations of von Neumann entropy for $d=3$ are done
by using \small{MATHEMATICA}.  For a given $\ka$ with $a=1$, we
determine $V_0$, the depth of the well and then the matching conditions
and the Fourier transform were used to obtain the entanglement spectra.
The \ee \ is then obtained by a numerical integration.  The results
are shown in a log-linear $S$ vs. $\ka$ plot in Fig
\ref{fig:1} which also shows the line obtained from Eq. \ref{eq:5} and
Eq.\ref{eq:threed}.  It shows that $S$ is negative for small $\ka$ and
that it has linear $\ln\, \ka$ dependence.  The approximations show
that the entropy is determined mainly by the outer part of the
wave-function.

\sfit

To generalize the result for any dimension we carried out the
calculation for general $d$.  The density matrix, solely from the
outer part, is expected to be of the form $f({\bf \tilde{q}})\sim
(1+\tilde{q}^2)^{-2}$ as in previous cases but then there is a
divergence problem for normalization for $d\geq 4$.  Since we want ${\rm Tr}\,
\hat{\rho} =1$, an ultraviolet cutoff is required.  This makes $a\ka$ an
important variable even in the limit $a\ka\to 0$.  The specialty of
$d=4$ is now evident.  }

The radial wave function $R(r)$ ($l=0$ state as the ground state) is, 
\begin{eqnarray} 
  \label{eq:6} 
  R(r) =
\begin{cases}
  A\ r^{\epsilon/2} J_{\mid \epsilon/2\mid} (k r)\quad {\rm for}\quad r<a\\ 
B \ r^{\epsilon/2} H_{\mid \epsilon/2\mid}^{(1)} (i \kappa r)\quad {\rm for}\quad r>a,
\end{cases}
\end{eqnarray} 
where $\epsilon=2-d$, $A,B$ determine the normalization and matching
of the inner and the outer solutions.  Here $J$ and $H^{(1)}$ are the
Bessel and the Hankel function of the first kind.  
{ The
  continuity of the wave function at $r=a$ gives
\bea
\label{eq:cont}
A\, J_{\left|\frac{2-d}{2}\right|}(k_c)&=& B\,(\ka
a)^{-\left|\frac{2-d}{2}\right|},
\eea
under  the condition $\ka\rightarrow 0$ and $ka\rightarrow k_c =\pi/2$. 
Eq.\ref{eq:expo} follows from Eq.\ref{eq:cont}, the matching of log
derivative and the Bessel function identities.
By using the normalization condition and Eq. \ref{eq:cont}, we 
get
\bea
 B &=& \begin{cases}
                                   \frac{\bka}{a}\hskip 1.2cm{\rm for}\ \ \
d<4\no\\
				    \frac{\bka^{\left|2-d\right|/2}}{a}\ \ \
{\rm for}\ \ \ d>4.
                                 \end{cases}
\eea
In the same $\kappa\rightarrow 0$ limit, with outer part dominance,
\beq
\phi(q=0)
\approx B\ka^{-\frac{2+d}{2}},
\eeq
which gives
\bea
\hat{\rho}({\bf q}) = |\phi({\bf q})|^2 &=& B^2\ka^{-(2+d)}\,f({\bf \tilde{q}})\no\\
&\approx& \begin{cases}
  \ka^{-d}\ \ \   {\rm for}\ \ d<4 \\
\ka^{-4} a^{d-4}\ \ \ \ {\rm for}\ \ d>4.
    \end{cases}
\eea
So the  von Neumann entropy is of the form Eq. \eqref{eq:5} with $P=4$ for
$d>4$.
}

In terms of the deviation from the critical point, the entropy is
\begin{equation} 
  \label{eq:8} 
  S=  \frac{d}{z(\Psi-1)} \ln\! \mid\! \!u-u_c\!\!\mid,\quad  {\rm for}\ \Psi<2. 
\end{equation} 
For the case in hand, $\Psi=d/2$.  The form of Eq.  \ref{eq:8} brings
out the universal behavior of the entropy and has validity for
potentials different from Eq. \ref{eq:1}, like e.g.  scale-free
$1/r^2$ potential\cite{smlong}.  All the details of the interaction
go in the universal exponents $z$ and $\Psi$.  The entropy diverges
at the critical point and, {\it is negative}.

\section{DNA connection}
We show the connection of the quantum entanglement entropy to the 
entropy of bubbles in DNA melting.  Under an imaginary time 
transformation, the path integral formulation of the quantum problem 
is analogous to a classical statistical mechanical system of polymers 
used in the context of melting of DNA\cite{fish,smb,fisher,doi}. 
 
 Let us consider a DNA whose two strands are two Gaussian
  polymers in $d$-dimensions and index the points (monomers) by the
  contour length $z$ measured from one end.  The native base pairing
  of a DNA requires that a monomer at index $z$ on one strand
  interacts with a point on the other strand with the same index $z$.
  This is the Poland-Scheraga type model\cite{fisher} for DNA melting.
  By using one extra coordinate for the sequence or the length of the
  polymers, we get directed polymers in $d+1$ dimensions like paths in
  path integrals, as shown in Fig \ref{fig:bubb}.  In this
  representation the base pairing interaction maps onto the same time
  interaction of the quantum system, time playing the role of the base
  pair index.  The DNA partition function as a sum over all polymer
  configurations is equivalent to the sum over all paths in quantum
  mechanics.  The DNA Boltzmann factor $\exp(-\beta H)$ with $\beta$ as the
  inverse temperature and $H$ the Hamiltonian for two chains of
  elastic constants $K_j$ as
\begin{equation}
{\beta H}\!\!=\!\!\!\!
  \int_0^N\!\! \left[\sum_{j=1,2}\!\!\frac{K_j}{2}\! \left(\frac{\partial
      \textbf{r}_j(z)}{\partial z}\right)^2\!\! +\!
  V(\textbf{r}_1(z)-\textbf{r}_2(z))\right]\!\! dz,
\end{equation}
corresponds to the factor $\exp(i{\cal S}/\hbar)$ for path integrals
with ${\cal S}$ the classical action of two interacting particles
under $z\to it $.  This makes the Green function or the propagator
${\cal G}(x_1,x_2,\tau|x_1^{\prime},x_2^{\prime},0)$ equivalent to the
partition function $Z(x_1,x_2,N|x_1^{\prime},x_2^{\prime},0),\ (N\to i
\tau)$.  Here $x_j, x_j^{\prime}$ are the coordinates of the $j$-th
strand end-points at $0$ and at length $N$.  The free energy per unit
length of DNA for $N\to\infty$ is the ground state energy of the
quantum problem.  

  The short range base-pairing potential can be taken to be
  a contact potential or a well of Eq. \eqref{eq:1}.  Then the picture
  of return of the quantum particles within the range of interaction
  after excursions outside the well gives the equivalent picture of
  polymers with broken base pairs having excursion away from binding
  and eventually coming back to the well to form pairs.  This
  excursion swells the polymer and creates bubbles along the length of
  the DNA.  Thermal energy opens up bubbles in the bound state of DNA.
  The entropy of a bubble of length $N$ is determined by the reunion
  partition function of two polymers starting together and reuniting
  again at $N$, which for large $N$, has the form $\Omega(N)\!\!=\!\!
  N^{-\Psi}e^{N \sigma_0}$, or the entropy 
\beq
\label{eq:13} 
S\equiv \ln \Omega(N) = N \sigma_0 - \Psi \ln N, 
\eeq 
in units of the
Boltzmann constant $k_B=1$.  Eq.  \ref{eq:13} shows that $\sigma_0$ is
the bubble entropy per unit length that survives in the thermodynamic
limit.  However, the power law $N$-dependence which gives the negative
sub-extensive part of the entropy is essential for the transition and
also for the bound state.  The reunion exponent $\Psi$ determines the
universality class of the binding-unbinding transition and there is a
melting transition if and only if $\Psi>1$.  See Ref. \cite{fish} for
details.

The one-dimensionality of the chains requires an alternating
arrangement of bound regions and bubbles as in Fig. \ref{fig:bubb}.
The arrangement allows one to write the partition function, after
Laplace transform with respect to the length (i.e. in the grand
canonical ensemble)\cite{fish}, as
\begin{eqnarray} 
  \label{eq:10} 
G(x,y;s) &=& G_{{\rm o}}(x;s)G(0,s)G_{{\rm o}}(y;s)\no\\
           &=&  \frac{G_{{\rm o}}(x;s)G_{{\rm o}}(y;s) G^{\rm B}(s,u)}{1- G^{\rm U}(s,\sigma_0)G^{\rm B}(s,u)}.
\end{eqnarray} 
Here $x\equiv \{x_1,x_2\}, y=\{x_1^{\prime},x_2^{\prime}\}$, $G_{{\rm o}}$ is
the Laplace transformed partition function of two polymers tied at one
end and open at the other, called the survival partition function, and
$G(0,s)$ is the total partition function with two ends bound.  In $G_{{\rm o}}$, 
the tied point is to be integrated over keeping the set
$x$ or $y$ fixed.  $G(0,s)$ can be written as a sum of a geometrical
series (see Fig. \ref{fig:bubb}) involving the partition functions of
the bound parts and the bubbles, $G^{\rm B}(s,u)$ and $G^{\rm U}(s,\sigma_0)$.
The free energy comes from the singularity of $G(x,y,s)$ which is
either $s=\sigma_0\equiv 0$ or at $s=s_0$ for which
\begin{equation} 
  \label{eq:12} 
 G^{\rm U}(s,\sigma_0)G^{\rm B}(s,u) = 1, 
\end{equation} 
with $\sigma_0=0$, $s_0$ satisfies Eq. \eqref{eq:expo}.

Near the nontrivial singularity, a pole at $s=s_0$, the form of
$G(x,y,s)$ resembles the Green function in the energy eigenfunction
expansion as
\begin{equation} 
  \label{eq:11} 
\frac{\langle y|\psi\rangle\langle \psi|x \rangle}{E-E_0} , 
\end{equation} 
with ground state dominance.  From the equivalence between DNA model
and the quantum problem, we identify the density matrix as
$$\rho(x,y)\sim G_{{\rm o}}(x;s_0)G_{{\rm o}}(y;s_0)/ G^{\rm U}(s_0),$$ 
so that the entanglement entropy would behave like \hbox{
$S\sim \ln G^{\rm U}(s_0,\sigma_0)$}.  By
using Gaussian distributions for Gaussian polymers (i.e. random
walkers), one recovers Eq. \ref{eq:dens}.  

To get the behaviour of $\ln G^{\rm U}$, we employ a finite-size
scaling analysis.  The phase transition in the polymeric system occurs
in the $N\to\infty$ limit so that a finite $N$ acts as a finite size
scale both for DNA and in the quantum problem.  The finite size
scaling variable is $N /\xi_{\perp}^z$ so that the entanglement
entropy is proportional to $ -z \ln \xi_{\perp} \sim
\frac{1}{\Psi-1}\; \ln \mid u - u_c\mid$ (see Eq. \ref{eq:8}).  The
difference in the amplitude occurs because of the different
normalization used for polymers and quantum problems.  The point to
note is that the entanglement entropy in the quantum problem comes from
the universal non-extensive part of the entropy of the bubbles.  Since
the full entanglement spectrum is known, it is also possible to
compute the Renyi entropy\cite{jphysa}.  We recover in the appropriate limit the
result quoted in Eq.  \ref{eq:5}.  In the DNA interpretation, the
Renyi entropy would come from many circular single strands (replicas)
pairing with a large single strand, resembling the rolling circle
replication of viruses.  Details will be discussed elsewhere.

\section{Discussion}
A negative entropy is counter-intuitive when one has the third law of
thermodynamics in the back of one's mind, though exceptions are known; e.g.
negative entropy is found for perfect gases at low temperatures or as
a corollary of the classical equipartition theorem.  One can see the
same feature by writing the reduced density matrix in terms of an
entanglement Hamiltonian, $\rho \propto \exp(-\beta H_{{\rm ent}})$, in a
form reminiscent of a Boltzmann factor. 
{The diagonal form in Eq.
\ref{eq:dens} shows 
\begin{equation}
  \label{eq:21}
  \beta H_{{\rm ent}}= 2 \ln (1+q^2/\ka^2)\approx 2 q^2/\ka^2, \ ({\rm for\
    small} \ q),
\end{equation}
which is like a classical $d$-dimensional oscillator in $q$-space,
with $\kappa^2$ as the effective temperature.  A direct calculation or
use of the classical equipartition theorem now tells us that the
entropy has $d\ \ln\ka$ behaviour as in Eq. \ref{eq:5}.}  We believe
this to be a generic feature whenever the entanglement Hamiltonian is
gapless.  Another way to see this emergence of $\ln \kappa$ in entropy
is to compare with the DNA problem.  The equivalent classical DNA
model also has a negative diverging part of entropy but that
sub-extensive part vanishes in the thermodynamic limit of the entropy
per unit length. In the quantum case, the equivalent limit has no such
advantage in finding the entropy because demanding extensivity in time
direction is meaningless. Hence the negatively diverging term is
inevitable near criticality.

In this paper we show that the quantum entanglement entropy near the
bound-unbound transition of two interacting particles comes out to be
negative, and it diverges at the QCP. Using the equivalent classical
statistical mechanical system of DNA near the melting transition we show
that the negativity of the entanglement entropy is a necessity and is
essential for the phase transition. The coefficient of the logarithmic
term contains the information of the interaction and the
universal behaviour of the phase transition.  The coefficient is
shown to be related to the reunion exponent of vicious walkers.  This
is the first time in the context of quantum entanglement that the
negative entropy is found by explicit calculation. We argue that this
log divergence in the quantum case and the sub-extensive part in the
DNA problem are linked by finite size scaling near the critical point.
From the renormalization group (RG) approach for the DNA melting
problem\cite{smb,smlong}, one may infer that the entanglement entropy increases along the RG flow, since the critical point corresponds to
the unstable fixed point.  It has been argued recently that
entanglement can be used to produce negative entropy\cite{rio}.  The
information theoretical meaning of the negative entropy in our case is
not very clear.  Our speculation is that the negative entropy is the
norm, not an exception near a quantum binding-unbinding transition. We
feel signatures of negative entropy might be detectable in cold atoms
where interactions can be tuned to the unitarity limit.  If one can
harness the negative entropy, one may cool a system or a computer and
possibly may overcome the obstacle to circuit miniaturization.


\end{document}